# A simulation sandbox to compare fixed-route, semi-flexible transit, and on-demand microtransit system designs

Gyugeun Yoon, Joseph Y. J. Chow[*], Srushti Rath
C2SMART University Transportation Center, Tandon School of Engineering, New York University, Brooklyn, NY, USA
[*]Corresponding author email: joseph.chow@nyu.edu

**Abstract**
With advances in emerging technologies, options for operating public transit services have broadened from conventional fixed-route service through semi-flexible service to on-demand microtransit. Nevertheless, guidelines for deciding between these services remain limited in the real implementation. An open-source simulation sandbox is developed that can compare state-of-the-practice methods for evaluating between the different types of public transit operations. For the case of the semi-flexible service, the Mobility Allowance Shuttle Transit (MAST) system is extended to include passenger deviations. A case study demonstrates the sandbox to evaluate and existing B63 bus route in Brooklyn, NY and compares its performance with the four other system designs spanning across the three service types for three different demand scenarios.

**Keywords**: public transport, semi-flexible transit, on-demand microtransit, Mobility-as-a-Service

# 1. Introduction

**1.1. Motivation**
Planning and design of public transit systems have become more complex with the emergence of new mobility services due to innovations in Information and Communications Technologies (ICTs) and the Internet of Things (IoT) (see Chow, 2018). Such decisions may now consider options between fixed route transit service or on-demand transit service mixed with other modes, and these decisions are all compounded by considerations of automation and electrification (WEF, 2019) as well as operation by public or private operators.

As a result of these advances, options for public transit provision include a range of services from conventional fixed-route transit to more on-demand, door-to-door "microtransit" service, and either in isolation or in combination with other services within a "Mobility-as-a-Service" (MaaS) platform. In such a platform, multiple services are available under a common gateway or platform to support travelers (Hensher, 2017). While Via, a shared ride microtransit provider, covers a certain area in cities where it operates vehicles (Via, 2019), transit agencies like New York City MTA tend to enhance the connectivity in cities by spanning out over the region (TRAVIC, 2020).

**Figure 1** provides a classification of modes available in MaaS platforms, which suggests a broad array of options for designing service solutions for travelers. A key take-away from this figure is that, despite the broad array of options, they mostly anchor around transit use. For example, active modes are often linked to public modes as part of multimodal trips (see Chow and Djavadian, 2015). The multimodality extends to shared modes as well, primarily in the first-last mile context (Ma et al., 2019a). They reinforce the importance of the public transit mode in the successful deployment of MaaS.

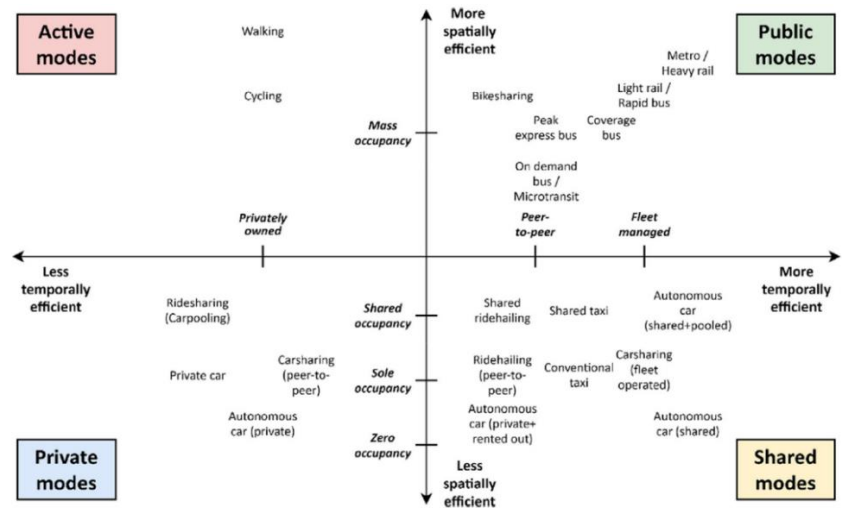

**Figure 1.** Different modes from Mobility-as-a-Service (source: Wong et al., 2020).

The success of such technological deployments has varied. Such microtransit operators as Bridj (Woodward et al., 2017), Kutsuplus (Haglund et al., 2019), and Ford Chariot (Korosec, 2019a) have all had to shut down operations. In the U.S., such operational challenges have led to the Federal Transit Administration (FTA) conducting pilots with programs like the Mobility-on-Demand (MOD) Sandbox Program to better understand the operations (GAO, 2018).

While the ICT technological advances have put the spotlight of transit innovations on recent years, the operation of Demand Responsive Transit (DRT) extend much further in history. For example, even from the late 1970s, the existence of a demand density threshold over which fixed-route service operates better than DRT was well-known (Systan, 1980). More recent insights (Daganzo and Ouyang, 2019) comparing fixed route transit with different types of on-demand service (microtransit, taxis) show the minimum door-to-door passenger travel time varies with fleet size for different levels of demand density for a stylized example. The result shows those factors affect the preference for various travel options in different ways, impacting the ranges, "sweet spots", where a mode is preferred over other modes regarding the passenger travel time.

**1.2. Research objective**

Clearly, successful MaaS deployment depends on making full use of the spectrum of transit operations, which can range from full fixed-route services, through flexible services, to full MOD-style microtransit services serving passengers door-to-door or virtual stop to virtual stop (Hazan et al., 2019). However, which types of services work well for a certain setting? How can a practitioner evaluate the service needs of a study with regards to different types of public transit services to support a MaaS ecosystem? To effectively compare different operating strategies within one study area without the costs and risks of full deployment, a simulation-based tool is needed.

We propose such a simulation sandbox that can be adapted to any study area to compare the operations of three classes (not mutually exclusive as hybrids between them also exist) of transit operations: fixed route transit, flexible-route transit, and on-demand transit. The term sandbox refers to a consistent framework for trying out different operating services so that their performances can be compared. The goal is to provide a transportation professional working in a public agency or a mobility provider with tools to help analyze and compare these services as societies progress toward a MaaS setting. The proposed simulation sandbox architecture is described in detail and its application is demonstrated in a case study using a common data set (a Brooklyn bus route, B63).

Accordingly, the remainder of this study consists of three main parts. Section 2 breaks down a review into each of the three categories of transit operations and the tools used to evaluate their performance. Section 3 describes the proposed simulation sandbox design including each system type, key design variables, and possible scenarios which can be implemented in the proposed sandbox. Section 4 provides a case study of an open-source tool and common data set, both made publicly available, for researchers and practitioners to assess their own study areas, and Section 5 concludes.

## 2. Literature review
**2.1. Classes of public transit operation**
Three classes of public transit operation are considered representing a range from most rigid (fixed route) to least (on-demand microtransit).

*2.1.1. Fixed-route transit service*

Certain vehicle technologies within this class generally require rigid routes through railway infrastructure (with exceptions, e.g. Cats and Haverkamp, 2018). Using a more rigid operational policy allows operators to attain much higher passenger flow capacities and, therefore, is more suited for high demand density populations as shown in Vuchic (1981).

Fixed-route transit service operational planning can be divided into several functions (Ceder, 2016): network route design, timetable development, vehicle scheduling, and crew scheduling.

Network design determines the structure and service of the network, including determination of routes and stops. Timetabling cements these route-level decisions with frequencies or headways along with a public timetable. Vehicle scheduling assigns the fleet to the timetables while crew scheduling assigns drivers and other staff to the fleet operations.

Planning routes and frequencies is a strategic level function. Reviews of transit network design models and algorithms can be found in Guihaire and Hao (2008) and more broadly in Farahani et al. (2013). A core part of network route design is the line planning, which determines the set of routes to operate and their frequencies. The line planning is NP-Hard in computational complexity (see Schöbel and Scholl, 2006). Early efforts to formulate and solve it (Hasselström, 1982; van Nes et al., 1988) handled fairly small instances. A related function within network route design is route generation to obtain the set of candidate routes. Route construction heuristics (e.g. Ceder and Wilson, 1986) have been used for this purpose.

For more specific designs at a route level, continuous approximation models have been proposed to design lines (Byrne, 1975; Newell, 1979), frequency (Newell, 1971; Mohring, 1972), and stop spacing (Vuchic and Newell, 1968; Wirasinghe and Ghoneim, 1981; Tirachini, 2014). In the frequency setting methods, the optimal frequency is shown to be proportional to the square root of the demand density. Furthermore, Chen et al. (2018) incorporated short-turn strategy to the local route service design to minimize total system cost. Fielbaum et al. (2016) characterized four different line structures using a common graph structure: direct lines, exclusive lines, hub-and-spoke, and feeder-trunk. Hub-and-spoke and feeder-trunk lines are particularly effective in addressing first-last mile problems to serve lower demand density areas like suburbs, and are an active area being targeted for MaaS integration with transit services (Ma et al., 2019a).

Metaheuristics have also been an alternative solution to the transit route network design problem, and Iliopoulou et al. (2019) reviewed literature that applied them and categorized studies into three types: single-solution-based, population-based, and hybrid metaheuristics, following Gendreau and Potvin (2005). While single-solution-based metaheuristics repetitively improve a single candidate solution, population-based metaheuristics examine the population which consists of candidate solutions. Hybrid methods are combinations of metaheuristics from both groups.

Meanwhile, tactical planning and operational control, lower levels of the transit route planning, involves other design elements. Timetabling, an example of tactical planning, involves setting the schedules at which a vehicle will service each stop (Ceder, 1987). The problem is complicated further when considering route synchronization (see Bookbinder and Désilets, 1992, and Ceder et al., 2001). Emerging data collection technologies to help with timetabling including automatic fare collection data (Yap et al., 2019) and autonomous shuttle bus service (Cao and Ceder, 2019).

### 2.1.2. *Semi-flexible transit*

In sparsely distributed demand regions, conventional fixed-route transit service is costly to serve due to the first-last mile problem (Chang and Schonfeld, 1991a). Researchers have proposed different transportation operations improve the degree of flexibility of fixed-route services. One such alternative is having DRT. DRT is the primary form of service in many rural and sparsely populated areas (FTA, 2018), and is the second largest transit service type in the US with 27,000 vehicles operating during peak service and $55M vehicle-revenue-hours.

The concept of DRT has existed since the 1970s and evolved with emerging technology such that there are now many different terms out there. Shaheen et al. (2015, 2016a) define a term, microtransit, to encompass the broad range of privately owned and operated shared transportation system that can offer fixed routes and schedules, as well as semi-flexible routes and on-demand scheduling. Volinski (2019) further expands on this definition for microtransit to include any shared public or private sector transportation services that offer fixed or dynamically allocated routes and schedules in response to individual or aggregate consumer demand, using smaller vehicles and capitalizing on widespread mobile GPS and internet connectivity. The only constant in these definitions is a responsiveness to demand. Accordingly, we distinguish two categories within this broad range of microtransit: a "semi-flexible transit" which maintains some degree of fixed route, and "on-demand microtransit" which corresponds strictly to door-to-door (or virtual stop to virtual stop) on-demand service.

As such, we examine flexible transit service provision across a spectrum from fully customized "on-demand microtransit" to "semi-flexible transit". On-demand microtransit provides door-to-door or stop-to-stop service to users without a fixed route. With semi-flexible route transit, on the other hand, there remain some stops of the fixed route that vehicles must drop by, known as checkpoints. A vehicle can deviate from a baseline connecting checkpoints and pick up or drop off passengers at optional stops called "virtual bus stops" to allow the service to vary to some degree. Rosenbloom (1996) surveyed 40 transit agencies and found that most of them had some flexible-route transit to remove or reduce the need to provide mandated complementary paratransit service, which can be much more expensive to operate. Koffman (2004) categorizes semi-flexible transit service into six main types: route deviation, point deviation, demand-responsive connector, request stops, flexible-route segments, and zone route. Errico et al. (2013) present an overview of the semi-flexible service classification and survey of methodological issues. Potts et al. (2010) surveyed transit agencies in North America and found that route deviation is the most common form of system (63.9%), whereas zone route accounts for 32.9%, request stops 30.9%, Demand-responsive connector service (DRC) 30.5%, flexible-route segments 19.5% and point deviation 16%.

Among the six types, "route deviation" has been studied the most by transportation researchers. Daganzo (1984a) first examined these "checkpoint dial-a-ride" systems and found that there's a relatively small window where they perform best: with higher demand densities they converge toward a fixed route service and at lower densities the service is outmatched by on-demand service. Quadrifoglio et al. (2007) presents a more state of the art version of route deviation system design called Mobility Allowance Shuttle Transit (MAST). In this system, a budget is designated on each route such that a single vehicle may only choose to deviate to a passenger request if it falls within the budget. Qiu et al. (2014) presented a flexible-route service design corresponding to "point deviation" in Koffman (2004).

Aside from the six types, Chen and Nie (2017) propose an alternative design that paired services with both fixed route and semi-flexible route as a hybrid, where their frequencies would be dependent on the demand density.

### 2.1.3. On-demand microtransit

The third operational class, and second within the microtransit category, is on-demand microtransit. This is the most flexible service type that is typically more costly to operate than the other service alternatives, but they can reduce costs when demand density is sufficiently low.

On-demand microtransit started as offline DRT to solve the Dial-a-Ride Problem (DARP). The first DARP models were proposed by Wilson (1967). The optimization of a service problem like DARP has been found to be computationally intractable (NP-hard) because it is a generalized case of a traveling salesman problem (TSP) which is also NP-hard (Papadimitriou and Steiglitz, 1977). As a result, evaluation of different routing-based service designs is often done using continuous approximation models for scalability.

For example, Beardwood et al. (1959) studied the approximate distances to serve a tour through multiple randomly generated points on a plane as a TSP and derived an asymptotic closed form expression. Stein (1978) proved that the length of an optimal tour to serve $n$ passengers with randomly determined pickup and drop-off locations (as a TSP with pickups and drop-offs, i.e. TSPPD) approaches a certain bound as $n \to \infty$ when ignoring wait and access time, for both single bus and multi-bus fleets. Daganzo et al. (1977) derived a many-to-one DRT system and a many-to-many DRT system (Daganzo, 1978) under different operating policies, using Beardwood's approximation as a starting point. The latter is studied as a queueing network.

A generalization of DARP, called Integrated Dial-a-Ride (IDARP), was developed by Häll et al. (2009) to integrate on-demand microtransit with fixed route transit as a first/last mile connector. In IDARP, passengers may change mode at transfer points to a fixed route service. The IDARP shares similarity in many aspects with the pickup and delivery problem with transshipments and the DARP with transfers (Cortes et al., 2010; Rais et al., 2014; Masson et al., 2013). The formulation is based on a directed graph formulation of the DARP (Cordeau, 2006) with an expansion that schedules both vehicle and customer itineraries.

Contrary to the static DARP and its variants, real-time operations of an online or dynamic DARP is not aware of passenger requests in advance. Dynamic routing has a long history from the late 1970s (Psaraftis, 1980; Psaraftis, 1995; Madsen et al., 1995; Agatz et al., 2011; Hosni et al., 2014). Real-time operating policies are divided into myopic and non-myopic policies: whereas a myopic policy only considers the decision at the time, a non-myopic policy anticipates cumulative costs from future decisions and outcomes over a time horizon using lookahead or other types of approximation.

One example variant of a myopic policy involves constraining the service to consider hubs and single transfers. Cortés and Jayakrishnan (2002) proposed a real time routing service called "High Coverage Point to Point Transit" (HCPPT) where each transit hub is designed for a group/cluster of such cells. The design strictly eliminates more than one transfer for any passenger and significantly decreases waiting time. This work was generalized by Jung and Jayakrishnan (2011). Another variant uses queueing theory to approximate the lookahead costs for making routing decisions, as exemplified by Pavone et al. (2010) (queueing network), Hyytiä et al. (2012) (online DARP), Sayarshad and Chow (2015, 2017) (online DARP with queue tolling, and relocation).

The use of real time routing with queueing and connection with public transit network as an online IDARP system design was proposed by Ma et al. (2019a). Non-myopic methods based on queueing (e.g. Sayarshad and Chow, 2015, 2017) were incorporated into the system design. The algorithm for the system design was tested for travel demand in Long Island to compare against the costs of operating an on-demand microtransit service.

Another important development to on-demand microtransit is the consideration of "meeting points" or virtual bus stops. Instead of picking up and dropping passengers off at their stated locations, the system would consider assigning them to common meeting points that may be a few blocks away. Stiglic et al. (2015) showed that such a system can improve matching rate and lead to mileage savings.

A subset of research is conducted on feeder systems to address the "last mile problem" in transit (see Chow and Djavadian, 2015). Chang and Schonfeld (1991a,b) first compared the relative advantages of fixed-route and DRT systems as feeder services and identified a similar demand density threshold under which DRT is more cost effective. They showed that for smaller service areas, higher express speeds, lower in-vehicle times, or higher access and wait times, flexible bus system becomes more advantageous. Quadrifoglio and Li (2009) proposed a method to identify a critical density for incorporating feeder buses.

In 2018 alone at least 24 agencies debuted microtransit pilots, but this considerable interest still begs the question of what role these services play and how they meet the mission of transit agencies (Lazo, 2018; Schaller, 2018). In most transit agencies, the worst performing fixed-route bus lines bottom out at around 10 passenger trips per vehicle hour (Walker, 2018b). Meanwhile, the best purely demand-response systems at best achieve 4 pickups per hour. Those which exceed that level to reach up to 7 or 8 do so by basing the service on a flex route pattern or anchoring one end of the service at a transit hub.

The interaction between shared on-demand mobility services and high-capacity public transit plays an important role. In recent years, the International Transport Forum (ITF) at the Organization for Economic Co-operation and Development (OECD) has conducted several simulation studies to investigate the impact that the shared on-demand mobility services would have on replacing other forms of transport such as traffic congestion, air pollution etc. (OECD/ITF, 2015, 2016, 2017a,b,c).

### 2.2. Synthesis of simulation-based methods for the three classes of transit operations

There are three main methods of evaluating different transit options: continuous approximation methods, computational methods using mathematical programming, and simulation-based methods. Analytical methods based on continuous approximation have been used since 1970s to compare fixed route and demand-responsive services, and later also for flexible-route services. The advantages of these methods are that they are simple to use, have closed form expressions that support sensitivity analysis, but the disadvantages are that they require simple assumptions about the underlying system and homogeneous demand patterns that make it difficult to map to real study areas. They are suited for research questions that can be answered using stylized examples. Computational methods may be more customizable to consider design variables and can characterize optimal performance but may not scale very well and require simplified assumptions about the decision process. Simulation allows rules to be generated in the greatest detail but the results of one simulation are not transferable and not as useful for sensitivity analysis.

Simulation-based evaluation of transit designs do exist in the literature (Cortés et al., 2005). Due to complexity of these systems, Djavadian and Chow (2017a,b) proposed an agent-based simulation to evaluate the market equilibrium of a broad range of flexible transit services through dynamics of day-to-day adjustment. MATSim, which is based on this day-to-day adjustment process, has been used to evaluate transit designs or as part of a simulation-based transit optimization process (Kaddoura et al., 2015; Nnene et al., 2019; Manser et al., 2020; He et al., 2021; Ma and Chow, 2021). They have also been used in evaluating demand-responsive transit (Narayan et al., 2017; Cich et al., 2017) and MaaS environment (Becker et al., 2020). Custom simulations have been used to evaluate fixed and flexible services. Ronald et al. (2015) provide a review of such studies. They then conducted a simulation comparison between on-demand microtransit and fixed route transit (Navidi et al., 2017), but did not include semi-flexible transit. Estrada et al. (2021) compared on-demand microtransit with taxi systems. Another recent study looked at simulations to evaluate MOD systems (Markov et al., 2021). However, these types of multi-agent simulations are designed for comprehensive travel demand modeling that captures dynamic traffic, agent activity scheduling and mode/route choice behavior. On the other hand, they do not capture detailed transit system design variables and do not have any semi-flexible route service considerations. There are empirical studies that have compared different transit designs (like the high-performance buses in Barcelona, see Estrada et al., 2012), as well as empirical evaluations of on-demand microtransit: for example, Alonso-González et al. (2018) studied a microtransit last mile feeder in the Netherlands; Shen et al. (2018) studied the integration of shared automated first-mile service in Singapore; Mendes et al. (2017) compared a fixed route light rail system with shared automated microtransit; while Haglund et al. (2019) conducted an ex post evaluation of the Kutsuplus system in Helsinki. To the best of our knowledge, there has not been any empirical or simulation study comparing between fixed route, semi-flexible, and on-demand transit.

Specific models for different transit operation policies can be established as introduced in following paragraphs.

First, fixed-route transit systems can be operated with vehicles traveling between stops located along unchangeable routes. Their strategic planning significantly affects the accessibility of transit services while tactical planning and operational control focuses on adjusting system elements without significantly impacting service demand. Since the purpose of a simulation sandbox is to compare users' responses to different system operation types, it focuses on strategic planning alternatives and excludes timetabling and vehicle scheduling. Assuming that the route structure is kept unchanged, it adjusts the number of stops and service frequency to optimize the total system cost. The details of optimizing two factors are described in Section 3.2.1; note, however, that the details are only provided for describing how a scenario was obtained, and these variables themselves become parameters within the simulation.

Second, flexible-route transit systems can implement the same route structure as the fixed-route system but have fewer stops. The most significant difference comes from higher flexibility in the vehicle travel between stops, allowing deviation from the baseline. It should share some common attributes with fixed-route service. The simulation sandbox implements a design for semi-flexible transit service among the variety of examples reviewed. The modeled service should be distinguished from other system classes because of its intermediate position between those. Considering this, the MAST system proposed by Quadrifoglio is chosen as the most appropriate design due to its intermediate structure between fixed-route system and on-demand microtransit, locating some fixed stops along the base route and allowing some flexibility of route deviation within the service area. The detailed design is illustrated in Section 3.2.2.

Third, on-demand microtransit system can have the highest flexibility with vehicles freely traveling passenger pickup and drop-off points regardless of any existing fixed route. Studies have investigated different aspects of them, improving the system performance from the perspective of either user's side, operators', or both. However, some advances are hardly applicable simultaneously because it not only complicates the problems to solve but also creates conflicts between objectives. Instead, the proposed simulation sandbox incorporates a basic on-demand microtransit service without developed extensions to provide clearer comparison results among different service types, leaving the potential to attach extensions as "modules." Based on these ideas, the direction and detailed design of the proposed sandbox are illustrated in the following section.

## 3. Proposed simulation sandbox design

A simulation sandbox is developed in this study to compare three alternative transit operation service system designs.

## 3.1. Simulation parameters

The simulation sandbox should be able to accommodate different system parameters and environmental factors and allow for consistent comparison of the three operating service system designs: a fixed-route line, a semi-flexible route service line (which we will call "flexible-route" service for simplicity), and on-demand microtransit. Parameters allow the modeler to fit the sandbox to different operations. These parameters are divided into three groups: simulation parameters that impact the simulation mechanism like time steps, system design variables like fleet size and vehicle capacity, and those reflecting operating conditions specific to the study area like passenger arrival rates, vehicle speeds, and passenger values of time.

*3.1.1. Simulation parameters*

The simulation is designed as a "discrete-time simulation" with a simulation length divided over discrete time steps. Each time step, vehicle states are updated according to the operating plan. For fixed-routes systems, the state is simply the vehicle location and passenger assignment. For flexible-route systems, the state also includes whether a vehicle deviates to serve a virtual stop. For on-demand microtransit, the state includes the sequence of passengers being served. Passenger states are also updated: waiting to be assigned to a vehicle, walking to a stop, waiting for vehicle, on-board a vehicle, egressing from vehicle stop to the destination. The simulation length and time step can be adjusted.

To prevent users from encountering the system without vehicles fully dispatched, the simulation sets warm up time $t_{wu}$ to locate vehicles along the route. Fixed-route and flexible-route systems require $t_{wu}$ at least twice longer than one-way cycle time to allow the first dispatched vehicle return to departed terminal, making sure vehicles be located along the route in advance. Otherwise, it will generate false results caused by temporal discrepancy between vehicle and passenger. For instance, a passenger should experience excessively long wait time if appearing in the region much earlier than the first vehicle arrival.

*3.1.2. Scenario parameters*

A common geographical boundary determines the service region and affects the system coverage and demand level. The service region is set into a rectangular shape of length ($L$) and width ($W$), which can be mapped from any shape route and corresponding catchments. The demand data should be collected within a service region according to defined boundary. Passenger arrival rate ($\lambda$), the number of passengers per unit time, defines the level of travel demand within the service region. The sandbox should accept artificially generated inputs as well.

The simulation takes inputs for weights of passenger travel time elements – in-vehicle time ($\gamma_v$), wait time ($\gamma_w$), and access time ($\gamma_a$). By multiplying them to each time element and adding them up, systems can compare the total passenger cost of candidate routes considering the unique values of time for passengers in the study area. For access time, walking speed ($v_w$) is necessary to calculate the access time, and the system sets a maximum walking distance ($\zeta_a$) to cover the effective catchment of a route or service.

Although average vehicle running speed ($v_o$) can be a feature of vehicle specification, it also can indicate congestion level that operators cannot control. For example, $v_o$ during peak hours should be lower than that in non-peak hours in urban areas.

*3.1.3. System design parameters*

Vehicle specification affects the capacity of the system. Vehicle capacity ($K$) limits the maximum number of passengers onboard the vehicle at any time. Fleet size ($V$) refers to the number of available vehicles in a system. Higher $v_o$ leads to faster passenger trips and lower vehicle relocation time.

The number of stops ($S$) in a fixed-route system and that of checkpoints ($S_c$) in a flexible-route system are key factors that affect the accessibility and mobility of systems. For example, larger $S$ or $S_c$ reduces average walking distance from/to stops, but vehicles should stop more frequently resulting in longer dwell times and stopping delay for passengers. On the other hand, the number of depots ($S_d$) determines the initial vehicle location in on-demand microtransit system and impacts the repositioning costs of idle vehicles. Average dwell time ($t_d$) defines the time that vehicles stay at stops to provide passenger with sufficient time to board or alight.

The average service frequency per hour ($f$) in fixed-route and flexible-route services affects the average headway ($h$) and wait time, key factors of both operators' and users' cost. $h$ should be the inverse of $f$ ($h = 1/f$) when both are observed for the same length of period. One-way cycle time ($t_c$) is the required time to finish a one-way trip between two terminals. Although it is a simple sum of vehicle running and stopping time in a fixed-route system, it needs additional time budgeted for deviations in a flexible-route system.

Lastly, maximum detour time rate ($\zeta_d$) in flexible-route and on-demand microtransit systems prevents overly long routes for a passenger caused by too many insertions between origin and destination. Maximum wait time ($\zeta_w$) is an upper bound on passenger wait time, ensuring passengers being served within the time threshold. Otherwise, systems reject pickup requests to let them find other modes (they leave the system as unserved). Maximum backtracking distance ($\zeta_b$) limits the distance traveled in the opposite direction by a flexible-route service, as described in Section 3.2.2. Maximum deviation distance defines the furthest distance from the checkpoint route that vehicles can reach in a flexible-route system but is usually equivalent to a half of service region width. **Table 1** summarizes the classification of parameters.

## 3.2. Transit operation system designs

This section delineates the three system designs, distinguishes the parameters specific to each design, and makes them adjustable in the simulation sandbox.

*3.2.1. Fixed-route*

Fixed-route operation considers a single transit line served between two terminals located on opposite ends of the rectangle with a distance of $L$. A fleet of buses departs from one terminal and head to the other, picking up and dropping off passengers at fixed stops along the fixed route.

For a fixed-route system, the modeler may choose to manually input the design parameters or let the simulation select the service frequency and stop spacing to minimize total cost. Total system cost consisting of users' and operator's cost is one of the most significant factors when considering the transit service performance measures (Desaulniers and Hickman, 2007). According to Tirachini (2014), estimating the total cost per hour $C_t$ which consists of operator cost per hour $C_o$ and user cost per hour $C_u$ is possible by using parameters as shown in Eq. (1) – (4).

**Table 1.** Classified parameters.

| Simulation parameters | | | |
|---|---|---|---|
| **Notation** | **Parameter** | **Notation** | **Parameter** |
| - | Simulation length | - | Time step |
| $t_{wu}$ | Warm up time | - | - |
| **Scenario parameters** | | | |
| **Notation** | **Parameter** | **Notation** | **Parameter** |
| $L$ | Route length | $v_o$ | Average vehicle running speed |
| $W$ | Service area width | $\gamma_v$ | Weight for passenger in-vehicle time |
| $\lambda$ | Passenger arrival rate (passenger/unit time) | $\gamma_w$ | Weight for passenger wait time |
| $v_w$ | Walking speed | $\gamma_a$ | Weight for passenger access time |
| $\zeta_a$ | Maximum walking distance | - | - |
| **System design parameters** | | | |
| **Notation** | **Parameter** | **Notation** | **Parameter** |
| $K$ | Vehicle capacity | $f$ | Service frequency |
| $V$ | Fleet size | $t_c$ | One-way cycle time |
| $S$ | Number of stops | $\zeta_d$ | Max. detour time rate |
| $S_c$ | Number of checkpoints | $\zeta_w$ | Max. wait time |
| $S_d$ | Number of depots | $\zeta_b$ | Max. backtracking distance |
| $t_d$ | Average dwell time | - | - |

$$C_t = C_o + C_u \qquad (1)$$

$$C_o = cft_c \qquad (2)$$

$$C_u = P_a \frac{L}{2v_w S} N + P_w \frac{1}{2f} N + P_v \frac{l}{L} t_c N \qquad (3)$$

$$t_c = \frac{L}{v_0} + \frac{\beta N}{f} + St_s \qquad (4)$$

where $c$ (\$/bus-h) is a unit bus operating cost, $f$ (bus/h) is bus frequency, $t_c$ (h) is the bus cycle time, $P_a$ (\$/h) is the value of access time, $L$ (mi) is the line length, $v_w$ (mph) is the walking speed, $S$ is number of stops, $N$ (passenger/h) is passenger demand, $P_w$ is value of waiting time, $P_v$ is value of in-vehicle time, $l$ (mi) is average travel distance per passenger, $v_0$ (mph) is bus operating speed, $\beta$ (hr/passenger) is average boarding and alighting time per passenger, and $t_s$ (h) is stopping delay.

Eq. (1) indicates that $C_t$ consist of $C_o$ and $C_u$, which can be derived by Eqs. (2) – (4). $C_o$ in Eq. (2) is the product of $c$ and $ft_c$ which can be the required fleet size to maintain $f$ on the route with $t_c$. Eq. (3) shows that $C_u$ can be divided into three components: access time, waiting time, and in-vehicle time. Meanwhile, Eq. (4) derives $t_c$ by aggregating average vehicle running time, total boarding and alighting time, and total stopping delay.

Reforming Eq. (1) as a function of $S$ and $f$ will lead to a nonlinear relationship between total cost and the two variables. We can find exact solutions of $(S^*, f^*)$ to this problem using any conventional nonlinear, unconstrained optimization method (e.g. gradient descent, Newton's method) since it's convex, although analytical, closed form expressions would not be attainable. Once the optimization is externally conducted, its result can be manually input to the simulation sandbox as a different configuration.

Consequently, the simulation sandbox can consider different manually input system designs for public transit. For example, one can use a preset configuration for an existing route to evaluate its performance in the sandbox, or conversely they can determine $(S^*, f^*)$ from the optimization of the total system cost via discretized enumeration and input them as the design variables. Note that the proposed simulation sandbox is not designed as an optimization tool. One of the design scenarios evaluated is based on a line-level optimal $(S^*, f^*)$, but other stop spacings and frequencies can also be used as input. This allows users to explore their own optimization methodology which can reflect specific criteria including geographical constraints and local regulations. For example, irregularly spaced stops can also be evaluated in a scenario by setting the coordinates.

### 3.2.2. Flexible-route

Its operation includes some relaxations of such operational constraints as route structure, stops, or timetables and is closer to a door-to-door service. The proposed simulation sandbox adopts and extends the MAST model from Quadrifoglio et al. (2007). It can be viewed as a hybrid of fixed-route system and on-demand microtransit and represent the intermediate class between two. This system design keeps the main skeleton of a fixed route and requires vehicles to drop by checkpoints, a subset of stops from a fixed route. Namely, vehicles are operated along the route connecting checkpoints. Passengers can use the service by either accessing those checkpoints along the baseline or requesting pickup or drop-off at their locations, dynamically generating virtual stops.

To accommodate deviations, a timetable is built for them considering some "slack time" in addition to normal vehicle operation time. For example, if a vehicle can run between two stops in 20 minutes, a timetable can allow an additional 10 minutes and indicate the travel time difference as 30 minutes. In this case, 10 minutes are assigned to a slack time to be used for deviating the vehicle to serve users arriving at one of the virtual bus stops during its run.

The other distinctive aspect of this system is a threshold of backtracking. Since a vehicle is permitted to deviate a route at its discretion, it can also reverse direction to pick up a passenger who suddenly shows up. However, when the distance of going back becomes excessively long, passengers who are onboard or waiting for vehicle may experience delays, leading to a worse aggregated user cost. Limiting the total length of backtracking can prevent these side effects while not rejecting all trips that require routing vehicles in the opposite direction. Maximum backtracking distance indicates the upper bound of the sum of backtracking distance in any section between sequential checkpoints.

This study extends the MAST model by including access by walking. In the original version (Quadrifoglio et al., 2007), a request from a passenger is matched with an available vehicle with sufficient slack time, and its pickup and drop-off points are inserted to the route of the vehicle if both can meet within the maximum wait time. In contrast, the modified version allows passengers to walk to nearby temporary vehicle stops or deviated vehicle routes if they can be served within the predetermined threshold. If users cannot receive the service, they abandon the given system and search for other travel alternatives, showing different behaviors from the fixed-route system. According to preliminary analyses with specific system configuration, the extension can process 23.6-88.0% more requests compared to the original version while imposing longer weighted travel time and total vehicle mile traveled (VMT). **Table 2** summarizes major differences in performance measures.

**Table 2.** Performance measure changes in the extended MAST compared to the original MAST.

| $\lambda$ (pax/h) | $S_c$ | Total ridership | | Avg. weighted travel time (min) | | VMT (mile) | |
|---|---|---|---|---|---|---|---|
| | | MAST | Extended | MAST | Extended | MAST | Extended |
| 80 | 10 | 167 | 314 | 61.94 | 70.97 | 343.36 | 339.69 |
| | 20 | 254 | 314 | 61.87 | 69.46 | 328.00 | 391.08 |
| 200 | 10 | 368 | 666 | 57.73 | 73.48 | 364.03 | 349.10 |
| | 20 | 560 | 753 | 60.94 | 69.27 | 328.36 | 419.32 |
| 400 | 10 | 696 | 1192 | 58.92 | 77.09 | 387.49 | 362.93 |
| | 20 | 1127 | 1546 | 59.99 | 71.38 | 328.12 | 423.41 |

### 3.2.3. On-demand microtransit

On-demand microtransit service resembles current shared ride-hailing services, which dispatch vehicles and transport passengers from their origins to their destinations. The system does not order vehicles to comply fixed routes or drop by mandatory stops. Instead, when assigning a passenger to a vehicle the system also updates the vehicle state which includes the sequence of pickup and drop-off points of passengers assigned to it. These matches determine vehicles' trajectories, passengers' travel experiences, and system performances.

Due to the influence of passenger assignment, it is crucial to apply the most appropriate methodology to balance the quality of the output and the instantaneousness of the reaction. This type of problems can be categorized into a vehicle routing problem with pickup and delivery (VRPPD) which has been extensively applied to the transport of the disabled and elderly, sealift and airlift of cargo and troops, and pickup and delivery for overnight carriers or urban services (Desaulniers et al., 2002).

The proposed simulation sandbox implements a simple insertion heuristic to update assigned routes for accepted passengers by searching for an updated sequence with the shortest incremental increase in travel time. The prior assigned sequence is not reassigned with each new passenger. If a route consists of $n$ stops, for example, there are $n$ available places where a new stop can be located, including the one after the last stop. Since the system would insert a pair of origin and destination (OD) locations with the precedence constraint that the origin cannot be located after the destination, locating the origin at the $i$-th place will limit the number of available places for the destination to $n + 1 - i$. The total number of candidate routes after insertion is shown in Eq. (5). If either origin or destination is already included in a set of stops that the route covers, the number of cases being reviewed can be reduced.

$$\sum_{i=1}^{n}(n + 1 - i) = \frac{n(n + 1)}{2} \qquad (5)$$

In addition to finding the route with the minimum total travel time increment, the study adds a function to the insertion heuristic that can verify the existence of passengers with unacceptable trips. If an insertion causes a significant increase of trip time of existing passengers, that combination of OD locations should be avoided. Because the main components of trip time are wait time and in-vehicle travel time, the algorithm can apply two thresholds. While the maximum wait time is an upper limit of wait time acceptable to passengers, the maximum detour time rate is the highest ratio of expected in-vehicle time after an insertion was made. If a passenger should spend additional time on a route that exceeds the product of the maximum detour time rate and direct travel time or the required time between OD if they can be directly connected without any intermediate stops, the route is considered infeasible. These constraints prohibit the system from imposing excessive cost to some passengers in the form of excessive wait and in-vehicle time. Thus, if the algorithm succeeds to suggest a route for a vehicle, that should demand the lowest total travel time among available routes which satisfies these thresholds for all passengers including a newly accepted one. Nevertheless, routing cost increment can be examined with some updates in the current version to consider both users' and operator's sides simultaneously.

*3.2.4. Summary*

**Figure 2** describes the conceptual visualization of simulated transit operation policies, indicating differences in vehicle routing flexibility, number of stops to be mandatorily visited, and pattern of passengers (indicated by grey triangles) accessing each system design.

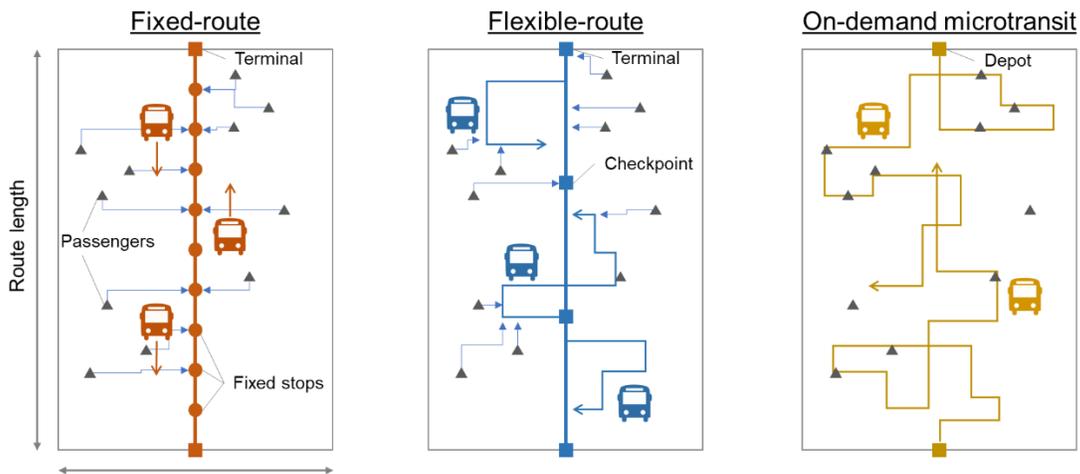

**Figure 2.** Concepts of simulated transit operation policies.

**Figure 3** briefly contrasts different flows of transit operation types. Fixed-route system does not consider the passenger rejection because it has no interactions with passengers in terms of the service feasibility. It can be assumed that passengers remain in the system until served. However, the other two systems consider possible routes through the insertion heuristics with requested pickup and drop-off points and may reject a passenger if no feasible routes exist to serve them. The flexible-route transit design allows reevaluation of passengers after they walk over to a new virtual bus stop location.

Some basic features are commonly embedded in those systems. For example, the simulation tracks the locations of vehicles moving with assumed running speed. It recognizes their arrivals when the coordinates of their current location equals to the next stop. After staying during the dwell time, vehicles depart the stop and head to the next one on their routes. Similar to vehicles, current locations of passengers are also monitored. Their typical trajectories are consisted of "actual origin – origin transit stop – destination transit stop – actual destination," walking to access or egress the transit service and riding vehicles between transit stops. They should wait at the origin transit stop if the vehicle has not arrived yet or already passed by.

**Algorithms 1** and **2** in the Appendices describe flexible-route and on-demand microtransit policies step by step and are implemented as shown in Figure 3. For fixed-route service, there is no dynamic policy in the simulation sandbox that needs to be implemented with an algorithm.

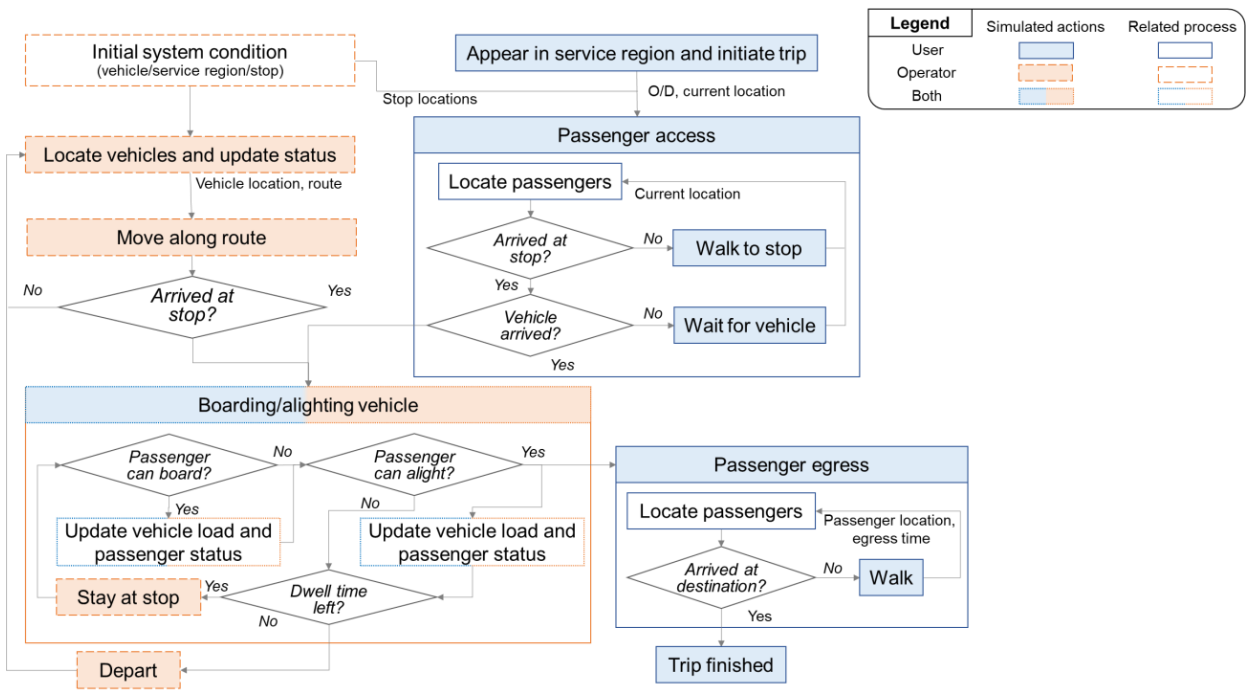

a) Fixed-route transit

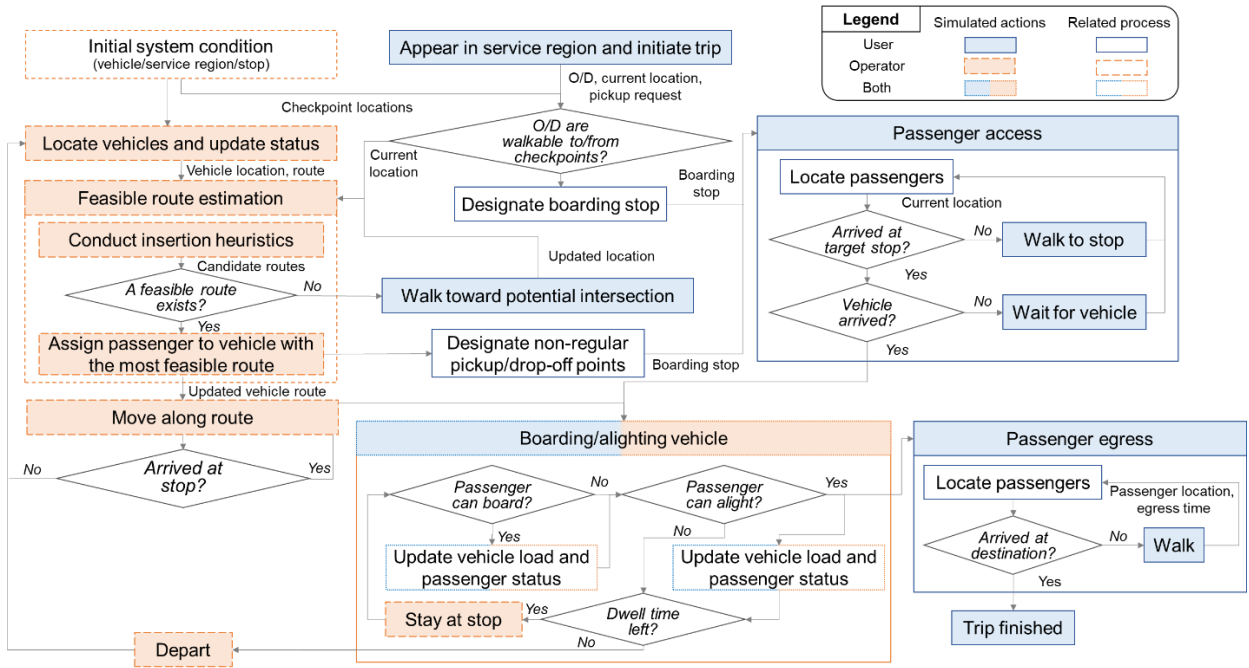

b) Flexible-route transit

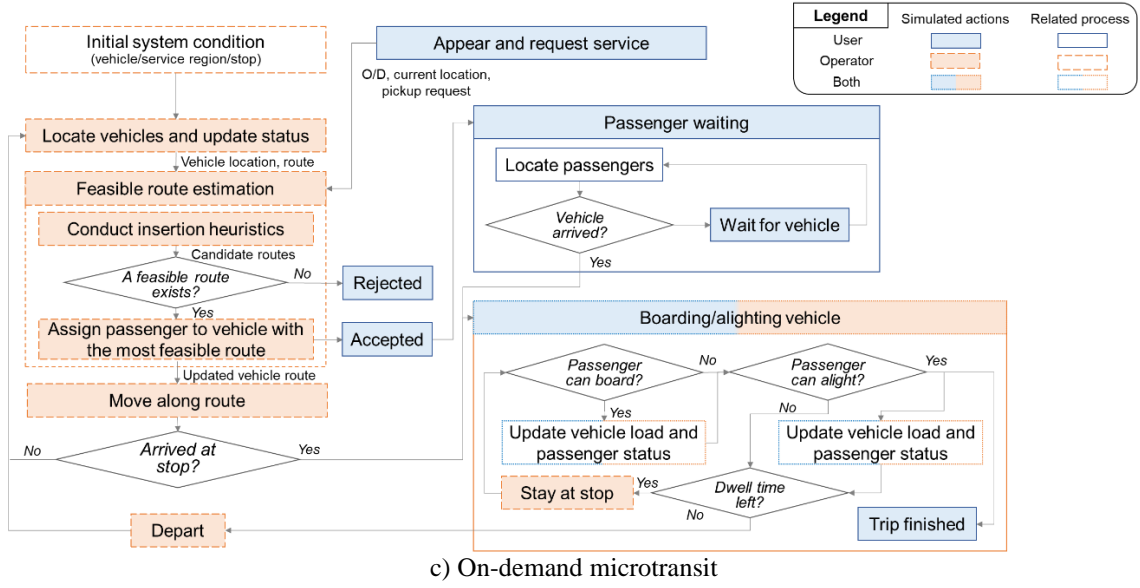

c) On-demand microtransit

**Figure 3.** Flow charts of transit operation systems for (a) fixed-route transit, (b) flexible-route transit (see Algorithm 1), and (c) on-demand microtransit (see Algorithm 2).

### 3.3. Testing system designs and scenarios

The purpose of the sandbox is to compare different transit operational system designs (fixed-route, flex-route, on-demand, and their design parameters) under different scenarios. Moreover, the sandbox can repeat multiple simulations with different design parameters. Consequently, the sandbox should conduct $N_{total}$ simulations as shown in Eq. (6), where $N_{vs,i}$ is the number of design variable sets being tested for system type $i \in \{fixed\ route, flexible-route, on-demand\}$ and $N_{dp}$ is the number of different scenarios.

$$N_{total} = N_{dp} \sum_i N_{vs,i} \qquad (6)$$

The sandbox is highly customizable. Instead of using random arrivals, it can be modified to receive an exogenous input arrival list which can be derived from spatial OD distributions observed from underlying activity zones (and has in fact been done so in Rath et al., 2022). Alternative route designs can then be considered; the locations of the ODs would simply need to be remapped relative to a new route design (e.g. an origin that starts off 1 km away from the original route at the 0.5 km mark along the alignment may become 0.5 km away from a second route design at the 0.8 km mark). More sophisticated trunk-feeder designs can also be considered by breaking up the analysis into one rectangular space per leg and adding up the performance measures, allowing for first- and last-mile analysis with the simulation of two-layered region: one of fixed-route trunk and the other for on-demand microtransit.

# 4. Common data and open-source simulation case study

An open-source simulation was developed from scratch that builds in the three types of transit operations to allow a decision-maker to compare them on any study area desired. This tool is demonstrated on a common data set in Brooklyn, NYC as a case study.

## 4.1. Simulation background

The Metropolitan Transportation Authority (MTA) operates 64 bus routes in Brooklyn in NYC, including Route B63, which connects northwestern and southwestern part of Brooklyn as shown in **Figure 4**. It has a length of 13.12 km (Stringer, 2017), covering 57 stops with 11,148 average weekday daily ridership in 2018 (MTA, 2021). The significant amount of ridership, the 16th largest among 53 routes in Brooklyn, indicates that it still provides accessibility to neighborhoods. While this route is currently operated in a fixed-route policy, we incorporate its study area into the simulation sandbox to compare performance under different demand density levels for different operating system designs.

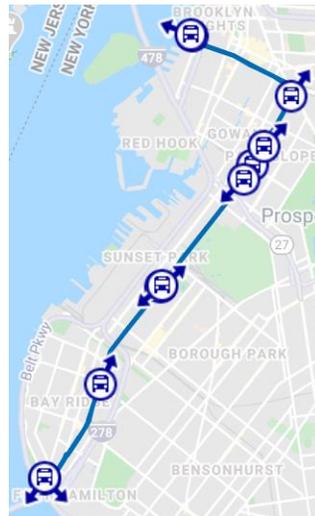

**Figure 4.** B63 route with vehicle locations in MTA Bus Time.

## 4.2. System design and scenarios

This case study simulates three scenarios based on different demand levels with five system designs as shown in **Figure 5**, where the existing fixed-route operation serves as one design and an optimized fixed-route design serves as another. The existing refers to the system observed in **Figure 4**.

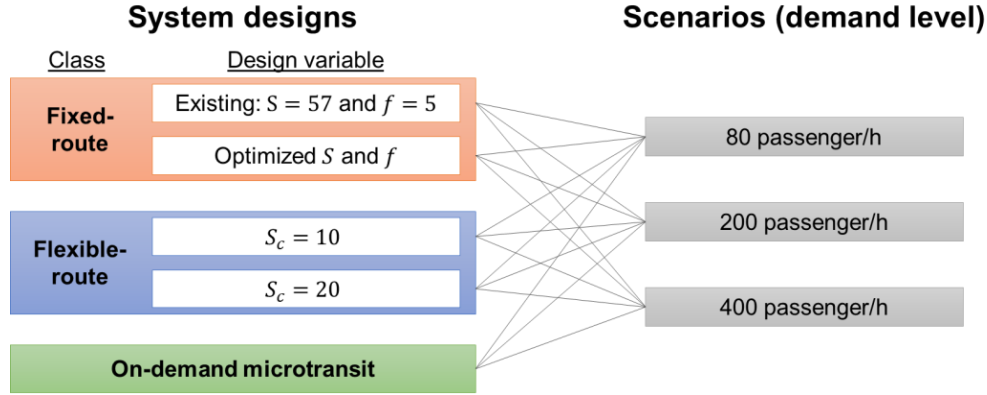

**Figure 5.** 15 combinations from 5 system designs and 3 scenarios.

**Table 3** indicates values of design variables, collected from various sources. Most are from Stringer (2017), bus route profiles in NYC. Assumed values do not have citations. Unless otherwise noted, all system designs share the value.

**Table 3.** Values of parameters

| Simulation parameters | | | |
|---|---|---|---|
| **Notation** | **Parameter** | **Notation** | **Parameter** |
| Simulation length | 4 hrs (14,400 secs) | Time step | 1 sec |
| $t_{wu}$ | $2t_c$ | | |
| **Scenario parameters** | | | |
| **Notation** | **Parameter** | **Notation** | **Parameter** |
| $L$ | 13.12 km (Stringer, 2017) | $v_o$ | 11.41 km/h (Stringer, 2017) |
| $W$ | 1.6 km | $\gamma_v$ | 1 (Wardman, 2004) |
| $\lambda$ | 80 [a], 200 [b], 400 [c] passenger/hr | $\gamma_w$ | 1.59 (Wardman, 2004) |
| $v_w$ | 5 km/h | $\gamma_a$ | 1.79 (Wardman, 2004) |
| $\zeta_a$ | 0.8 km (Zhao et al., 2003) | - | - |
| **System design parameters** | | | |
| **Notation** | **Parameter** | **Notation** | **Parameter** |
| $K$ | 85 [d1,d2] (MTA, 2019)/ 40 [e1,e2] / 20 [f] | $f$ | 5 [d1,e1,e2] (MTA, 2020b), 1.5 [d2,a], 2.4 [d2,b], 3.6 [d2,c] vehicle/h |
| $V$ | 15 [d1,d2] / 20 [e1,e2] / 40 [f] | $t_c$ | 88 [d1,d2] (Stringer, 2017) / 120 min [e1,e2] |
| $S$ | 57 [d1] (Stringer, 2017), 30 [d2,a], 33 [d2,b], 36 [d2,c] | $\zeta_d$ | 2 [f] |
| $S_c$ | 10 [e1] / 20 [e2] | $\zeta_w$ | 12 [e1,e2] / 30 min [f] |
| $S_d$ | 10 [f] | $\zeta_b$ | 0.4 km [e] |
| $t_d$ | 20 sec | - | - |

Note) [a]: scenario with 80 passenger/hr, [b]: scenario with 200 passenger/hr, [c]: scenario with 400 passenger/hr, [d1]: reference fixed-route system, [d2]: optimized fixed-route system, [e1]: flexible-route system with 10 checkpoints, [e2]: flexible-route system with 20 checkpoints, [f]: on-demand microtransit. Otherwise, values applied to all cases.

MATLAB, a commercial computer programming language, is used for coding the simulation. The simulation assumes several conditions and parameters of which some are introduced previously. The simulation covers four hours, equivalent to 14,400 time steps of one second.

While the peak hour passenger demand is approximately 800 passenger/hr, the case study evaluates lower demand levels 10%, 25%, and 50% of peak hour demand to observe the change of performance as the demand level varies (which can reflect less busy off-peak periods). ODs of artificial passengers are randomly generated and evenly distributed within the area, limiting ODs not to be connectable via walking. Their arrivals in the area are assumed to follow the Poisson process. All scenarios use the identical passenger dataset for different demand levels to prevent potential disturbances caused by the heterogeneity of demand during the comparison among policies. Furthermore, the behaviors of passengers are assumed not to be elastic regarding the provided service level. The only reason of their exclusion from the system is the violation of $\zeta_d$, $\zeta_w$, and $\zeta_b$, representing the threshold of expected service level.

### 4.3. Results

**Figure 6** is a sample illustration of vehicle trajectories following each system design for the same amount of time in the case study. The fixed route passes through the middle of the area, a semi-flexible route deviates from the fixed route if there is a non-regular point, and an on-demand microtransit service vehicle freely moves within the area. From captured trajectories and data, the simulation can aggregate some performance measures of both vehicles and passengers, total ridership, average weighted travel time, and total VMT, as shown in **Tables 4 – 6** and **Figure 7 – 9**. Total system cost is not directly compared due to the uncertainty of system design parameters for different operation policies. For example, the operated vehicle capacity is assumed to vary, causing the difference of unit operating vehicle cost. Instead, alternative measures provided in this section can allow the simulation users to roughly estimate total system cost based on previous practices and experiences.

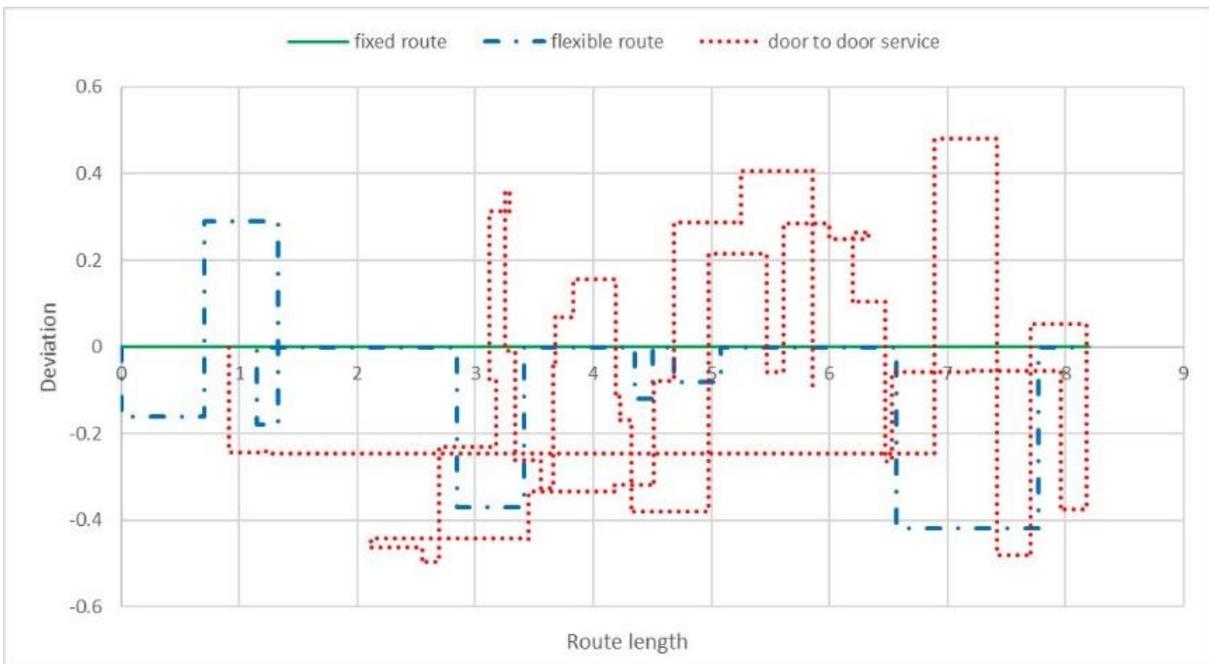

**Figure 6**. Sample of simulated vehicle trajectories.

**Table 4.** Simulated total ridership.

| $\lambda$ (pax/h) | Fixed (Existing) | Optimized Fixed | Flexible ($S_c$=20) | Flexible ($S_c$=10) | Microtransit |
|---|---|---|---|---|---|
| 80 | 333 | 331 | 314 | 314 | 318 |
| 200 | 791 | 791 | 753 | 666 | 586 |
| 400 | 1625 | 1622 | 1546 | 1192 | 766 |

**Table 5.** Simulated average weighted travel time (min).

| $\lambda$ (pax/h) | Fixed (Existing) | Optimized Fixed | Flexible ($S_c$=20) | Flexible ($S_c$=10) | Microtransit |
|---|---|---|---|---|---|
| 80 | 58.86 | 72.22 | 69.46 | 70.97 | 47.61 |
| 200 | 58.63 | 69.19 | 69.27 | 73.48 | 54.40 |
| 400 | 58.78 | 63.93 | 71.38 | 77.09 | 53.49 |

**Table 6.** Simulated total vehicle mileage (mi).

| $\lambda$ (pax/h) | Fixed (Existing) | Optimized Fixed | Flexible ($S_c$=20) | Flexible ($S_c$=10) | Microtransit |
|---|---|---|---|---|---|
| 80 | 335.72 | 137.73 | 339.69 | 391.08 | 772.10 |
| 200 | 335.72 | 160.71 | 349.10 | 419.32 | 1005.31 |
| 400 | 335.72 | 228.80 | 362.93 | 423.41 | 1072.54 |

Ridership increases as demand increases for all scenarios, but increments vary. While fixed-route scenarios achieve the ridership proportional to prevailing demand, others experience some deficits, meaning that their performances are less efficient than the fixed-route designs. For flexible-route designs, fewer checkpoints may lead to lower ridership as a smaller portion of demand can reach checkpoints and vehicles require more slack time to reach more non-regular points. On-demand microtransit system only serves nearly half of the ridership of fixed-route service leaving the other half unserved. The main reason of this number drop can be the complete connection between actual origin and destination of users eliminating all access and egress trips covered by users' feet.

Average weighted travel time is the shortest with on-demand microtransit service as it excludes passenger walking of which penalty is the highest. It also rejects user requests with estimated wait time longer than $\zeta_w$ and in-vehicle time exceeding $\zeta_d$. In flexible-route systems, the compliance of timetable with slack time can extend both wait and in-vehicle time, resulting in longer travel time than others. Numbers in on-demand microtransit and flexible-route transit increase according to the demand level because more insertions of pickup and drop-off points are made along vehicle routes. For fixed-route system, user-perceived travel time remain the same due to the unchanged number of stops and frequency keeping wait time and in-vehicle time constant. Meanwhile, the optimization of $S$ and $f$ imposes more travel cost to users since it tends to save operator's cost when demand level is low. Decreasing weighted travel time along the demand indicates "economy of scale" in fixed-route service.

Total VMT with on-demand microtransit service is the longest among scenarios, 2.3 to 3.2 times longer than the reference fixed-route service. This may indicate the trade-off between passenger travel time and VMT and. For fixed-route system design, the optimization seems to help reduce vehicle operation significantly, which causes longer travel time for passengers due to fewer stops and less frequency. Flexible-route designs yield slightly higher VMT than fixed-route designs. The influence of demand level is significant only in on-demand microtransit service while others maintain the similar level.

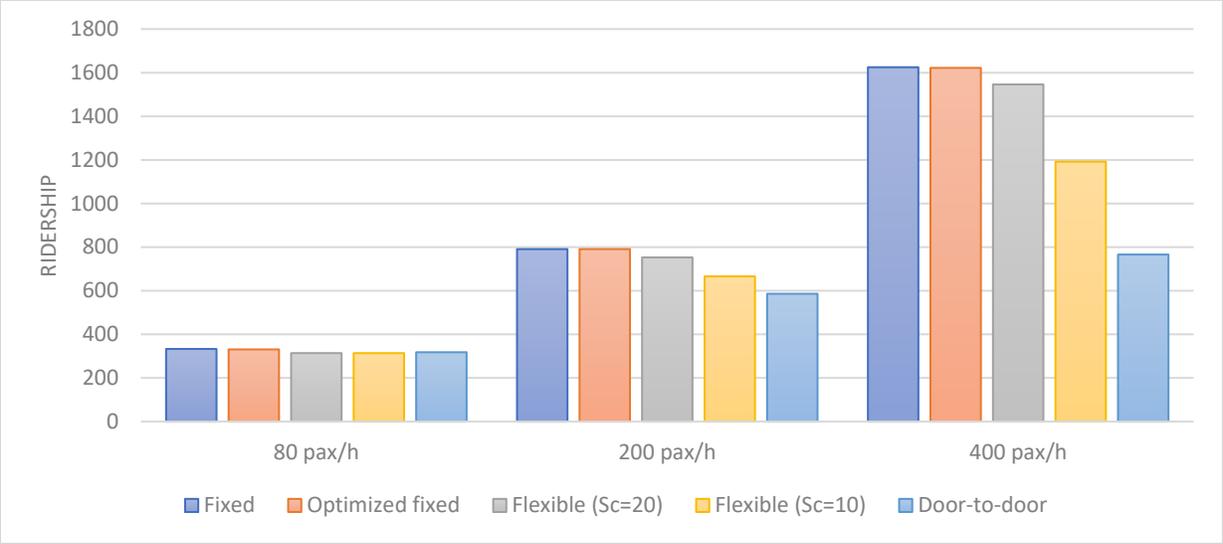
**Figure 7.** Simulated total ridership.

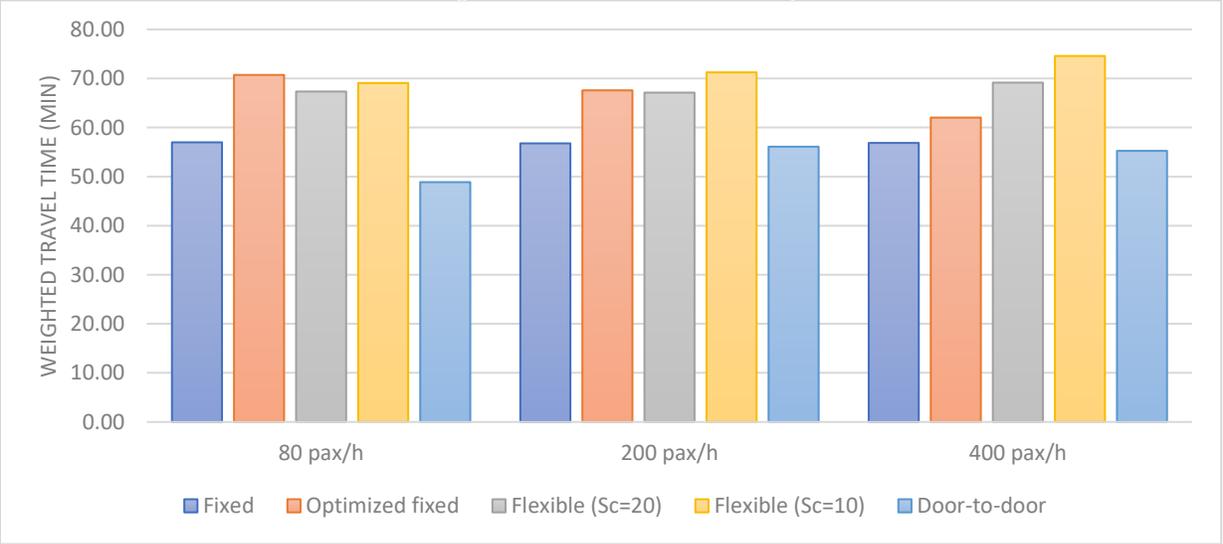
**Figure 8.** Simulated average weighted travel time.

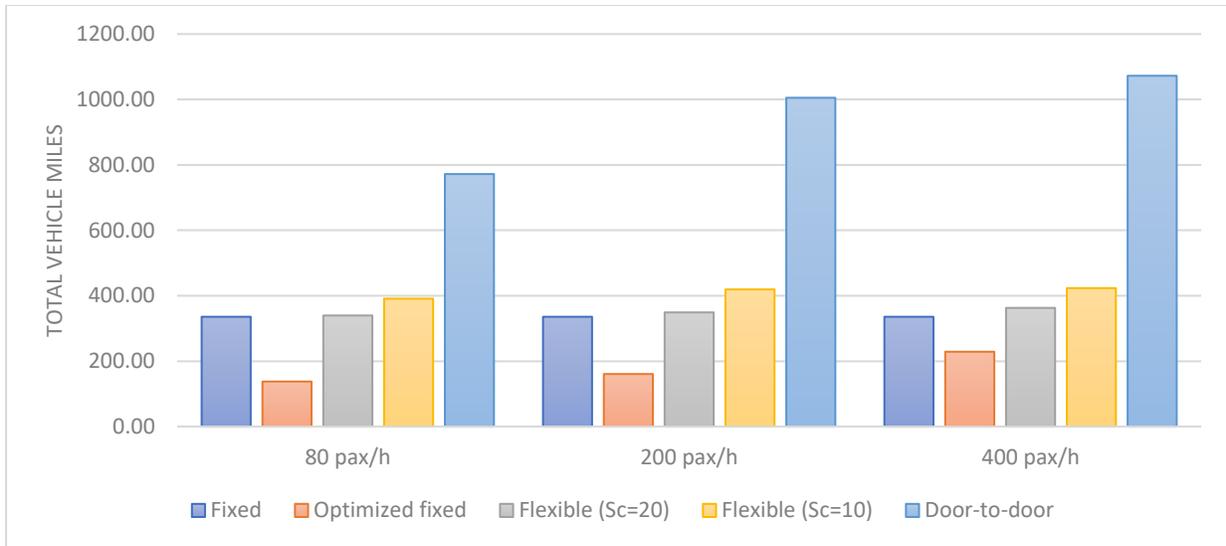

**Figure 9.** Simulated total vehicle mileage.

## 5. Conclusion

This study provides a new simulation sandbox for transit operations planning considering the range of fixed-route through semi-flexible transit to on-demand microtransit. The sandbox simulates different classes of transit operation to evaluate their performances under different design variables and scenarios. For the fixed-route service, manually input design variables are allowed as well as having the sandbox optimize frequency and stop spacing. For the semi-flexible route transit operation, a new extension of the MAST design from Quadrifoglio et al. (2007) is implemented to allow passengers to reposition to another location so that they can be fit into the deviation budget of the vehicle, which increases the acceptance rate. To the best of our knowledge, such a sandbox has not been developed in the literature.

A case study is presented to demonstrate this tool using a common dataset derived from the B63 bus route in Brooklyn, NYC. Three performance measures – total ridership, average weighted travel time, and total VMT – vary along different system designs and scenarios. First, fixed-route systems show the largest total ridership while others serve fewer passengers under the same demand pattern. Although the main reason can be the availability of rejection in flexible-route and on-demand microtransit service, it also indicates how efficient the system designs in the case study are. Second, on-demand microtransit service can provide the shortest average weighted travel time among system designs despite the decreasing gap as demand level elevates. This is common in systems with any flexibility. However, optimized fixed-route systems can reduce it by providing higher frequency when facing higher demand. Lastly, total VMT is the longest in on-demand microtransit system due to the exclusion of access/egress walks of users. The measure of optimized fixed-route system increases as it operates higher frequency. Other system designs maintain the similar level of VMT regardless of demand level.

This simulation sandbox can potentially be applied to all transit lines in the U.S. to output performance metrics throughout so that relationships can be established between different local built environments, their regulatory and institutional settings, and investment levels with performance metrics under different types of operating policies. For example, this can be done as a supplement to the data provided in the National Transit Database or as part of the Transit Equity

Dashboard from TransitCenter (Klumpenhouwer et al., 2021). In addition, this is a starting point for other efforts. Next steps include route redesign, first- and last-mile analysis, evaluation of multimodal system with multiple operators, transfer coordination, and consideration of vehicle electrification.

# Appendix

**Algorithm 1.** Extended MAST insertion with passenger walking and multiple vehicles for flexible-route policy

Input: $L, W, T, t_d, v_o, v_w, U, V, K, h, S_c, t_c, \zeta_a, \zeta_b, \zeta_w$

Initialization: Locate $S_c$ checkpoints evenly distributed along fixed route and define the specification of vehicle $i$ where $i \in V$ ($K$, $v_o$, initial routes in both direction (from Checkpoint 1 to $S_c$ and vice versa), travel time ($t_t = \frac{L}{S_c-1}\frac{1}{v_o}$) and slack time ($t_{slack} = \frac{t_c}{S_c-1} - t_t - t_d$) between checkpoints, and dispatch time ($\tau_{di} = (i-1)h$)). $i_{max} = 0$.

For $\tau = 1, 2, \cdots, T$ do
1. Dispatch vehicle $i$ when $\tau = \tau_{di}$, $i_{max} = i_{max} + 1$.
2. If there are services requests from passenger $j \in U$, go to the next step. Otherwise, go to Step 13.

For $i = 1$ to $i_{max}$ do
   [Direct service]
   3. Recall information of $i$ including existing route $r_i$ according to the identified direction of passenger $j$ travelling.
   4. Calculate expected wait and in-vehicle time of passengers assigned to $i$, $t_{slack}$ for each section between checkpoints, and performance measure.
   
   For $k_1 = 1$ to $|r_i|$ do
     For $k_2 = k_1 + 1$ to $|r_i| + 1$ do
       5. Insert $O_j$ at the $k_1$-th sequence of $r_i$ and $D_j$ at the $k_2$-th place to create $r_i^*$.
       6. Investigate the feasibility (the violation of $K$ $t_{slack}$, and $\zeta_b$). If feasible, calculate updated expected wait and in-vehicle time of passengers and performance measure. Otherwise, try the next set of $k_1$ and $k_2$.
     Next
   Next
   7. Choose $r_i^*$ with the minimum impact on performance measure as a feasible candidate route $r_{id}^*$. If there is no available route, $i$ cannot directly serve $j$.
   
   [$j$-walking]
   8. Identify segments $r_i$ which $j$ can access from $O_i$ and egress to $D_j$ within $\zeta_a$ and determine potential intersections $s_i$ where to approach (start/mid/end point).
   
   For $k_3 = 1$ to $|s_i|$ do
     9. Create $r_i'$ by inserting $k_3$-th intersection to $r_i$, and calculate the impact on expected wait and in-vehicle time of passengers and performance measure.
     10. Investigate the feasibility (the violation of $K$, $t_{slack}$, $\zeta_b$, and $\zeta_w$). If feasible, calculate updated expected wait and in-vehicle time of passengers and performance measure. Otherwise, try the next $k_3$.
   Next
   11. Choose $r_i'$ with the minimum impact on performance measure as a feasible candidate route $r_{iw}^*$. If there is no available route, $j$ cannot be assigned to $i$.
Next
12. Assign the passenger to either $r_{id}^*$ or $r_{iw}^*$ with the minimum impact of the performance measure. Update information of $i$ and $j$. If there is no available $i$, send $j$ to the rejected passenger set $U_r$.

[Resurrection of rejected passengers $j_r$]

13. If $|U_r|>0$, repeat from Step 3 to 12 only for $j_r$ who reach every 30 sec after their rejection except for ones added in this time step.

[Vehicle relocation]

For $i = 1$ to $|V|$ do

14. If vehicle is at a stop, determine whether a vehicle stays more or leaves a stop based on remaining dwell time. If there are additional passenger in this time step, process them.
15. If vehicle is moving, determine whether a vehicle keeps moving or arrives at a stop based on remaining distance. When arriving, process passengers waiting for being picked up or dropped off.

Next

16. If time step does not reach simulation period, go back to Step 1. Otherwise, go to Step 1.

Next

17. Aggregate simulation outputs.

---

**Algorithm 2.** Insertion heuristic for dispatching and routing on-demand vehicles

Input: $L, W, T, t_d, v_o, v_w, U, V, K, h, S_d, \zeta_w, \zeta_d$, vehicle distribution along depot $\mu_s$

Initialization: Locate $S_d$ depots evenly distributed along fixed route and define the specification of vehicle $i$ where $i \in V$ ($K, v_o$, empty routes,). The number of vehicles per depot $n_s$ is defined by the discrete distribution $\mu_s$ where $\sum_s n_s = V$.

For $\tau = 1, 2, \cdots, T$ do

1. If there are services requests from passenger $j \in U$, go to the next step. Otherwise, go to Step 8.

[Passenger assignment]

For $i = 1$ to $i_{max}$ do

2. Recall information of $i$ including existing route $r_i$.
3. Calculate expected wait and in-vehicle time of passengers assigned to $i$ and performance measure.

For $k_1 = 1$ to $|r_i|$ do
  For $k_2 = k_1 + 1$ to $|r_i| + 1$ do
  4. Insert $O_j$ at the $k_1$-th sequence of $r_i$ and $D_j$ at the $k_2$-th place to create $r_i'$.
  5. Investigate the feasibility (the violation of $K, \zeta_d$, and $\zeta_w$). If feasible, calculate updated expected wait and in-vehicle time of passengers and performance measure. Otherwise, try the next set of $k_1$ and $k_2$.
  Next
Next

6. Choose $r_i'$ as a feasible candidate route $r_i^*$ with the minimum impact on performance measure. If there is no available route, $i$ cannot serve $j$.

Next

7. Assign the passenger to $r_i^*$ with the minimum impact of the performance measure and update information of $i$ and $j$. If there is no available $i$, reject $j$.

[Vehicle relocation]

For $i = 1$ to $|V|$ do

8. If vehicle is staying at a stop, determine whether a vehicle stays more or leaves a stop based on remaining dwell time. If there are additional passenger in this time step, process them.
9. If vehicle is moving, determine whether a vehicle keeps moving or arrives at a stop based on remaining distance. When arriving, process passengers waiting for being picked up or dropped off.

Next

10. If time step does not reach simulation period, go back to Step 1. Otherwise, go to Step 1.

Next

11. Aggregate simulation outputs.

**Availability of data and materials**
The datasets generated and/or analyzed in the study are available in:
- Simulation sandbox code: https://github.com/BUILTNYU/FTA_TransitSystems
- Generated Brooklyn case study data set: https://doi.org/10.5281/zenodo.3672151

**Funding**
The authors were supported by an FTA grant NY-2019-069-01-00 and the C2SMART University Transportation Center (USDOT #69A3551747124).

**Acknowledgements**
Help from NYU MS student Patrick Scalise in preparing the literature review and NYU Abu Dhabi student Sara Alanis Saenz in preparing the case study data are appreciated. Professor Quadrifoglio shared the insertion heuristic code for his MAST algorithm with us which is much appreciated.

van Nes, R., Hamerslag, R., & Immers, L. H. (1988). The design of public transport networks. *Transportation Research Record* 1202, 74-83.

Via (2019), Via on-demand transit system, URL https://ridewithvia.com/ (accessed 10.24.2019).

Via (2018). Getting Microtransit Right, https://www.intelligenttransport.com/transport-whitepapers/68280/whitepaper-getting-microtransit-right/.

Volinski, J. (2019). TCRP Synthesis 141: Microtransit or General Public Demand–Response Transit Services: State of the Practice. Federal Transit Administration.

Vuchic, V.R. (1981). Urban public transportation: systems and technology. Prentice-Hall, Englewood Cliffs, N.J.

Vuchic, V. R., & Newell, G. F. (1968). Rapid transit interstation spacings for minimum travel time. *Transportation Science*, 2(4), 303-339.

Walker, J. (2018b). Is Microtransit a Sensible Transit Investment? Hum. Transit. URL https://humantransit.org/2018/02/is-microtransit-a-sensible-transit-investment.html (accessed 5.17.19).

Wardman, M. (2004). Public transport values of time. *Transport policy, 11*(4), 363-377.

WEF (2019). Shared, electric and automated mobility (SEAM) governance framework: prototype for North America and Europe. World Economic Forum White Paper, July 19th, https://www.weforum.org/whitepapers/shared-electric-and-automated-mobility-seam-governance-framework-prototype-for-north-america-and-europe.

Wilson, N. H. (1967). *Computer Aided Routing System* (Doctoral dissertation, Massachusetts Institute of Technology, Department of Civil Engineering).

Wilson, N.H., Sussman, J., Wong, H.-K., Higonnet, T. (1971). Scheduling algorithms for a dial-a-ride system. Massachusetts Institute of Technology. Urban Systems Laboratory.

Wirasinghe, S. C., & Ghoneim, N. S. (1981). Spacing of bus-stops for many to many travel demand. *Transportation Science*, 15(3), 210-221.

Wong, Y. Z., Hensher, D. A., & Mulley, C. (2020). Mobility as a service (MaaS): Charting a future context. *Transportation Research Part A: Policy and Practice*.

Woodward, C., Vaccaro, A., Gans, F. (2017). Bridj, local on-demand bus service, is shutting down. *Boston Globe*, April 30th.

Yap, M., Luo, D., Cats, O., van Oort, N., & Hoogendoorn, S. (2019). Where shall we sync? Clustering passenger flows to identify urban public transport hubs and their key synchronization priorities. Transportation Research Part C: Emerging Technologies, 98, 433-448.

Zhao, F., Chow, L. F., Li, M. T., Ubaka, I., & Gan, A. (2003). Forecasting transit walk accessibility: Regression model alternative to buffer method. *Transportation Research Record, 1835*(1), 34-41.
29